\documentclass[preprint,5p,twocolumn,sort&compress]{elsarticle}

\usepackage{amssymb}
\usepackage{hyperref}
\usepackage{amsmath}
\usepackage{graphicx}
\usepackage{color}

\newcommand{\beq}{\begin{equation}}
\newcommand{\eeq}{\end{equation}}

\journal{Physics Letters B}

\begin{document}

\begin{frontmatter}



\title{Hot New Early Dark Energy:\\ Towards a Unified Dark Sector of Neutrinos, Dark Energy and Dark Matter}


\author{Florian Niedermann}
\ead{florian.niedermann@su.se}

\address{Nordita, KTH Royal Institute of Technology and Stockholm University,\\
            Hannes Alfv\'ens v\"ag 12, 
            SE-106 91 Stockholm,
            Sweden}

\author{Martin S.~Sloth}
\ead{sloth@cp3.sdu.dk}
\address{CP$^3$-Origins, Center for Cosmology and Particle Physics Phenomenology \\ University of Southern Denmark, Campusvej 55, 5230 Odense M, Denmark}

\begin{abstract}
Hot new early dark energy describes a supercooled, first-order phase transition that takes place at sub-eV temperatures in the dark sector. It lowers the sound horizon, which provides a possible solution to the Hubble tension, and, at the same time, it can explain the neutrino masses through the inverse seesaw mechanism by making a set of sterile Majorana fermions massive. First, we argue that this scenario strengthens existing cosmological bounds on the heaviest neutrino mass. This, in turn,  constrains the dark sector temperature, which provides us in total with two falsifiable predictions. In a second step, we discuss the phenomenological consequences of embedding hot new early dark energy in a larger gauge group that is partially broken above the TeV scale. This novel theory, which could even be motivated independently of the Hubble tension, completes the high-energy corner of the inverse seesaw mechanism and explains the mass of a dark matter candidate that can be produced through gravitational interactions at high energies.
\end{abstract}



\begin{keyword}
New early dark energy \sep phase transition \sep neutrino mass generation \sep Hubble tension

\PACS 98.80.Cq \sep 98.80.-k \sep 98.80.Es


\end{keyword}

\end{frontmatter}

\section{Introduction}

The nature of dark energy (DE) and dark matter (DM)  are great open questions in cosmology, which, at the same time, provide evidence that the Standard Model (SM) of particle physics is incomplete. In fact, the SM has its own internal consistency problems, with the hierarchy problem and the question regarding the origin of neutrino masses being two of the most prominent ones. So far, we lack solid evidence of what new physics will resolve these questions, but this situation could be about to change. The Hubble tension, one of the hottest and most debated topics in cosmology at the moment~\cite{Freedman:2017yms,Verde:2019ivm,DiValentino:2020zio}, could be evidence of new physics at the eV scale operative before recombination~\cite{Verde:2016ccp,Bernal:2016gxb,Knox:2019rjx,Aylor_2019,Arendse:2019itb,Efstathiou:2021ocp}. Informed by the new eV physics required to solve the Hubble tension, we propose a unified framework for the dark sector, connecting an early DE component, neutrinos, and DM.

It has been known for a while now that one of the most promising avenues for resolving the Hubble tension in cosmology is an extra component of DE, which decays just before recombination. In the first proposal of this type, called early dark energy (EDE)~\cite{Poulin:2018dzj, Poulin:2018cxd,Poulin:2018cxd,Smith:2019ihp,Smith:2020rxx,Murgia:2020ryi,Poulin:2021bjr}, the decay of this extra component of DE happened just before recombination in a second-order rollover of an axion-like scalar field (also see~\cite{Lin:2019qug,Kaloper:2019lpl,Alexander:2019rsc,Hardy:2019apu,Sakstein:2019fmf,Berghaus:2019cls,1798362,Lin:2020jcb,CarrilloGonzalez:2020oac,Freese:2021rjq,Allali:2021azp,Sabla:2021nfy,Karwal:2021vpk,Vagnozzi:2021gjh,Gomez-Valent:2021cbe,Seto:2021xua,Moss:2021obd,Clark:2021hlo} for related work and \cite{Escudero:2019gvw,Kreisch:2019yzn,Park:2019ibn,Pandey:2019plg,Sekiguchi:2020teg,Jedamzik:2020krr,Escudero:2021rfi,Bansal:2021dfh,Aloni:2021eaq} for other early-time approaches). But without fine-tuning the axion potential the EDE component does not dilute fast enough after the transition to be in agreement with data. In addition, in order to agree with observations sensitive to the axion-like field perturbations, the potential must be flattened at high field values excluding a simple monomial form, and finally, the model cannot resolve the $\sigma_8$ tension of large-scale structure (LSS), although it does not make it worse either~\cite{DAmico:2020ods,Ivanov:2020ril,Hill:2020osr,Murgia:2020ryi,Smith:2020rxx}. On the other hand, a triggered first-order phase transition has the potential to naturally resolve these issues, as was shown in the new early dark energy (NEDE) proposal~\cite{Niedermann:2019olb,Niedermann:2020dwg,Niedermann:2020qbw}. In the NEDE model, a fast first-order phase transition is triggered by a scalar field that becomes light and starts moving in the potential before matter-radiation equality. While addressing the fine-tuning issues of the old EDE model, the NEDE model is also phenomenologically different at the perturbation level and has different predictions for CMB and LSS. A first combined fit to CMB, BAO, and SNe data alongside a local prior on $H_0$ revealed a $4 \sigma$ evidence for a non-vanishing fraction of NEDE  within the two-parameter model, bringing the tension down to the $2.5\, \sigma$ level~\cite{Niedermann:2019olb}. In the meanwhile, NEDE has been generalized through the inclusion of more free parameters, tested without a local prior on $H_0$, and compared with EDE using different data set combinations~\cite{Niedermann:2020dwg,Niedermann:2020qbw,Allali:2021azp,Schoneberg:2021qvd}. The upshot is that EDE and NEDE are equally strong phenomenological models, but differences are expected to arise with more precise CMB temperature and polarization data. In fact, the first hints of these differences have already emerged in a recent ACT analysis, although it remains to be seen how much they are driven by internal inconsistencies between Planck and ACT data~\cite{Poulin:2021bjr}. Beyond that, the idea of using a first-order phase transition during the CMB epoch has led to further interesting NEDE model-building efforts that, for example, consider a chain of tunneling events~\cite{Freese:2021rjq} or a confinement phase transition~\cite{Allali:2021azp}. 

Here, we will consider a recently proposed~\cite{long} version of NEDE where the first-order phase transition of a complex scalar field $\Psi$ is triggered by finite temperature corrections to its vacuum potential (hot NEDE) instead of an ultra-light scalar field (cold NEDE). In Sec.~\ref{sec:hotNEDE}, we will review the mechanism and demonstrate that a sizable fraction of NEDE can be obtained in the supercooled regime where the eV-scale transition temperature $T_d^*$ has dropped significantly below the critical temperature $T_c$ (at which the potential develops a new minimum for the first time). In particular, we will derive a bound on the self-coupling $\lambda$ that ensures that the field stays in thermal equilibrium with the dark sector plasma as required for the thermal field theory computation to be applicable.

Quite generically, the supercooled regime implies a vacuum expectation value (vev) for $\Psi$ which exceeds $T_d^* \sim \mathrm{eV}$. Incidentally, the inverse seesaw mechanism -- an elegant way to create the active neutrino masses~\cite{Abada:2014vea}  -- requires the existence of super-eV sterile neutrinos. It has therefore been argued that the NEDE phase transition gives mass to a set of sterile neutrinos $\nu_s$ through a Yukawa interaction, which, in turn, renders the active neutrinos massive~\cite{long}.\footnote{Although not as phenomenologically successful as NEDE, different attempts to relate the Hubble tension and the neutrino sector have been made in the past~\cite{Escudero:2019gvw,Escudero:2021rfi,Fernandez-Martinez:2021ypo,DiBari:2021dri}, most notably~\cite{Sakstein:2019fmf,CarrilloGonzalez:2020oac} explores the possibility that the EDE field is pushed up its potential when neutrinos become non-relativistic (see also \cite{DAmico:2018hgc} for a DE neutrino interaction). To our knowledge, our work is the first to use the NEDE phase transition for higgsing the neutrino sector.} As a new result, in Sec.~\ref{neutrino_masses}, we will derive a consistency relation between the phenomenological NEDE parameters and the mass of the heaviest active neutrino, which can be tested by fitting the model to cosmological datasets. It also implies a lower bound on the dark sector temperature, which is an equally falsifiable prediction and will inform our further model building.

At this level, hot NEDE is a low-energy framework compatible with different microphysical embeddings.  However, we believe it is important to showcase how these general ideas can be made more concrete in a specific particle physics model. To that end, we explore in Sec.~\ref{sec:DEW} a microphysical description of hot NEDE that was proposed recently in ~\cite{long} and is dubbed \textit{dark electroweak model} (DEW). It is invariant under a dark sector gauge group $\mathrm{SU(2)}_\mathrm{D} \times \mathrm{U(1)_\mathrm{Y_D}} $, which is broken down to a dark electromagnetism $\mathrm{U(1)_\mathrm{DEM}} $ close to the TeV scale. The NEDE scalar $\Psi$ is then identified as a the $\mathrm{U(1)_\mathrm{DEM}} $ neutral component of an $\mathrm{SU(2)}_\mathrm{D}$ triplet $(\Psi_1,\Psi_2,\Psi_3)^T$. This high-energy embedding achieves different things: It completes the high-energy corner of the inverse seesaw mechanism that requires a Dirac mixing between the sterile neutrinos and a set of right-handed neutrinos; it introduces an approximate global lepton symmetry $\mathrm{U(1)}_\mathrm{L}$ that is spontaneously broken in the hot NEDE phase transition and protects the neutrino masses against radiative corrections; it provides a concrete framework for calculating thermal corrections to the vacuum potential; and finally, it accommodates the mass generation of a super-TeV DM candidate that can be produced through gravitational interactions close to the Planck scale. Here, we will argue that it can be made compatible with the new phenomenological bounds on the dark sector temperature. Finally, in Sec.~\ref{sec:pheno}, we discuss the model's phenomenological signatures both in cosmology and particle physics experiments.

In short, this letter studies the phenomenology of hot NEDE when applied to both the neutrino mass problem and the Hubble tension. This is first done at low (sub-TeV) energies but then extended to a larger dark sector gauge theory valid at higher energies. In doing so, this work aims at providing a minimal and viable example of how the Hubble tension can guide our quest for a complete dark sector model.

\section{Hot NEDE Phase Transition} \label{sec:hotNEDE}

Hot NEDE introduces a non-vanishing temperature $T_d$ in the dark sector. Here, will assume that the hidden sector contains a new dark gauge group with gauge coupling parameter $g_\mathrm{NEDE}$, which is spontaneously broken when the NEDE field $\Psi$ acquires a non-vanishing vev as the dark temperature drops below $T_d^* \lesssim \mathrm{eV}$. Adopting the Abelian Higgs model as our working example and denoting the modulus of the complex scalar field $\Psi$ with  $\psi = \sqrt{2} |\Psi|$, the temperature-corrected potential takes the form~\cite{long} (although we expect the same structure to arise in other gauge theories)

\begin{multline}
\label{eq:effective_T_pot_low_T}
V(\psi;T_d) =  - \frac{1}{8} \,g_\mathrm{NEDE}^2T_\circ^2 \psi^2 +\frac{\lambda}{4}\psi^4 \\
+ 3 T_d^4 K\left( g_\mathrm{NEDE} \psi / T_d\right) \mathrm{e}^{-g_\mathrm{NEDE} \psi / T_d} + V_0(T_d)\,,
\end{multline}
where $V_0$ denotes $\psi$-independent contributions and $K(a)$ is a mildly varying function such that $0.1 < | K(a)| \lesssim 10 $ for $0 < a \equiv g_\mathrm{NEDE}\psi/T_d  < 30$. It admits the expansion $K(a) \mathrm{e}^{-a}\simeq - \pi^2/90 + a^2/24 + \ldots$ for small argument.
To be precise, in~\cite{long}, it was shown that, within that range, it can be approximated by (with an error smaller than $4\%$)
\begin{multline}\label{eq:K}
 K(a)= -0.1134 \, (1+a) - 0.0113 \, a^2 +\\
 4.32 \times 10^{-6} \ln{(a)}\,a^{3.58}+0.0038\,\mathrm{e}^{-a(a-1)} \,,
\end{multline}
which is compatible with the expansion $K(a) \mathrm{e}^{-a}\simeq - \pi^2/90 + a^2/24 + \ldots$ that can be derived analytically for small argument~\cite{Dolan:1973qd}.
Moreover, $T_\circ $ sets the temperature scale of the phase transition and is related to the vacuum mass $\mu$ through  $T_\circ = 2 \mu / g_\mathrm{NEDE}  $.
This potential is valid in the perturbative regime where~\cite{Arnold:1992rz,long} $\lambda \lesssim g_\mathrm{NEDE}^3$ and reproduces three decades old results of  the effective temperature-dependent potential in the limit where the gauge boson mass $m_A = g_\mathrm{NEDE} \psi$ is small compared to the temperature.\footnote{To be precise, we recover the result in \cite{Dine:1992wr} when identifying $8 D_\mathrm{there} \simeq (4 \pi E_\mathrm{there})^{2/3} =g_\mathrm{NEDE}^2$ and taking the high-temperature limit $g_\mathrm{NEDE} \psi / T_d \ll 1$. } 
However, for hot NEDE to work, we need the generalized form in \eqref{eq:effective_T_pot_low_T}, which is also valid for small temperatures and allows us to describe a situation where the vacuum energy dominates over the dark radiation plasma. This supercooled regime corresponds to $\gamma \lesssim 1$ where~\cite{long}
\begin{align}\label{eq:gamma}
\gamma = \frac{4 \pi \lambda}{g_\mathrm{NEDE}^4} \, .
\end{align}
We further note that for \eqref{eq:effective_T_pot_low_T} to be applicable, the NEDE field has to be in thermal equilibrium with the gauge sector. We will argue later that this imposes a lower bound on the self-coupling parameter $\lambda$ that can be easily satisfied.
Apart from $\gamma$ and $T_0$, the vacuum structure of the potential is conveniently characterized by
 \beq\label{eq:delta_large_mass}
\delta_{\mathrm{eff}}(T_d) = \pi \gamma \left(1-\frac{T_\circ^2}{T_d^2}\right) \, .
\eeq
To be specific, a metastable minimum at $\psi=0$ exists for $0 < \delta_\mathrm{eff}(T_d)\lesssim \pi \gamma$ {sufficiently small}. This can be seen by expanding $K(a)$ for small argument, {$K(a) \mathrm{e}^{-a}\simeq - \pi^2/90 + a^2/24$}, which results in a term quadratic in $\psi$ with positive coefficient {(corresponding to a local minimum at $\psi=0$)}. This expansion breaks down when the local maximum around  $ \psi \sim T_d / g_\mathrm{NEDE}$ is reached, and the exponential suppression kicks in. For even larger values of $\psi$, the true vacuum is found at $\psi \simeq v_\Psi  \equiv  \mu / \sqrt{\lambda}$. For $\delta_\mathrm{eff}(T_d) \to  \pi \gamma$, on the other hand, the second minimum disappears, and $\psi=0$ is the global minimum.  This corresponds to the initial state of the system at high temperatures. Then, as the dark sector cools, $\delta_\mathrm{eff}(T_d)$ decreases until the second minimum develops. The thermal transition becomes efficient when the percolation parameter  $p\sim \Gamma /H^4$ exceeds unity. Here, $H$ denotes the Hubble parameter, and we introduced the decay rate per volume~\cite{Linde:1981zj},
\begin{align}\label{Gamma_3}
\Gamma \sim T_d^4 \exp{\left( - S_3 / T_d\right)}\,,
\end{align}
where $S_3$ denotes the Euclidian action in the $\mathrm{O(3)}$-invariant case (this contribution can be shown to dominate over the $\mathrm{O(4)}$-invariant saddle-point, which is relevant for cold NEDE).  
Assuming supercooling ($\gamma \lesssim 1$) and taking the thick-wall limit ($\delta_\mathrm{eff} \ll \pi \gamma$), we have \cite{long} 
 \beq\label{S_3_large_mass}
\frac{S_{3}}{T_d} \sim  \frac{\sqrt{4 \pi}}{g_\mathrm{NEDE}^3 \gamma}  \delta_\mathrm{eff} ~.
\eeq
Analogous to cold NEDE~\cite{Niedermann:2020dwg}, we find that for an eV scale phase transition the percolation condition $p\sim 1$ is satisfied as $S_3/T_D^* \simeq 250$. This condition together with~\eqref{S_3_large_mass} then fixes
\begin{align}\label{delta_eff_star}
\delta^*_\mathrm{eff} \equiv  \delta_\mathrm{eff}(T_d^*) \sim 10 \gamma g_\mathrm{NEDE}^2\,,
\end{align}
which is compatible with the thick-wall approximation for a perturbative coupling with $g_\mathrm{NEDE} \lesssim 0.1$. 

We define NEDE as the false vacuum energy at $\psi=0$ normalized with respect to the true vacuum at $\psi=v_\Psi$. With this, the fraction of NEDE is
$f_\mathrm{NEDE} = -V(v_\Psi;T_d \to 0)/ \rho_\mathrm{tot}$, where $\rho_\mathrm{tot}(t_*)$ is the total energy density at the time of the phase transition, $t_*$ (henceforth an asterisk denotes a quantity's evaluation at $t_*$). We can further evaluate it through \eqref{eq:effective_T_pot_low_T} and \eqref{eq:delta_large_mass} as
\begin{align}
f_{\textrm{NEDE}} 	\simeq   \frac{\pi}{16} \frac{1}{\gamma}  \frac{T_d^{*4}}{\rho_\mathrm{tot}(t_*)} \quad\quad (\text{for}\,\,\gamma \lesssim 1\,\, \text{and} \,\, g_\mathrm{NEDE} \lesssim 0.1 )~.\end{align}
Assuming that $\rho_\mathrm{tot}$ is dominated by the visible sector radiation, we can invert the relation and express the relative dark sector temperature $\xi = T_d / T_\mathrm{vis}$ as a function of $f_\mathrm{NEDE}$ 
\begin{align}\label{xi_large_mass}
\xi^4_* \simeq 0.56 \times \gamma \left[\frac{f_{\textrm{NEDE}} /(1-f_{\textrm{NEDE}} )}{0.1}\right] \,,
\end{align}
or in absolute terms
\begin{align}\label{eq:T_d_star_large_mass}
T_d^{*4} \simeq (0.7 \mathrm{eV})^4 \gamma \left[\frac{f_{\textrm{NEDE}} /(1-f_{\textrm{NEDE}} )}{0.1}\right]\left[ \frac{1+z_*}{5000}\right]^4\,,
\end{align} 
where $z_*$ is the redshift at the time of the phase transition. The suggested values, $f_\mathrm{NEDE} = 0.1$ and $z_* = 5000$, correspond to phenomenologically interesting choices within the cold NEDE scenario~\cite{Niedermann:2020dwg}.
We observe that we can easily have a large fraction of NEDE with $f_\mathrm{NEDE} \simeq 0.1$ alongside a cold dark sector with $\xi < 0.5$, if we demand $\gamma \ll 1$ . This supercooled regime corresponds to $T_d^*/T_c \sim \sqrt{\gamma} \ll 1$ and equally implies that we have a strong first-order phase transition where the released vacuum energy dominates over the dark radiation plasma. In position space, the phase transition corresponds to the nucleation of bubbles of true vacuum that expand and eventually collide. The bubbles' typical sizes are bounded by the time $\bar{\beta}^{-1}$ it takes to convert all of space into the true vacuum. Using the definition $\bar{\beta} = \mathrm{d} (S_3 T_d^{-1})/\mathrm{d}t $, we derive $ H_*\bar{\beta}^{-1}  \sim  10^{-2} g_\mathrm{NEDE}^2$. In order to prevent bubbles from growing to cosmological scales, which would lead to large anisotropies in the CMB~\cite{Niedermann:2020dwg} and LSS~\cite{Freese:2021rjq}, we require   $H_*\bar{\beta}^{-1} < \mathcal{O}(1) \times 10^{-3}$, which is satisfied for a coupling parameter $g_\mathrm{NEDE} < 0.1$.  

Now, resolving the Hubble tension demands that NEDE leads to a  sizable, i.e.~$\mathcal{O}(10\, \%)$, but short energy injection into the cosmic fluid before recombination. The supercooled metastable vacuum before the transition achieves a quick build-up of such an injection that goes as $a^4$. 
After the phase transition, the system is described by a colliding bubble wall condensate of typical length scale $\bar{\beta}^{-1} \ll 1/H_*$. It is characterized by huge gradients and eventually dissipates into radiation. On large scales  $\gg \bar{\beta}^{-1}$, it can be described as a fluid with time-dependent equation of state $w_\mathrm{NEDE}(t)$. As the state will be initially dominated by small-scale shear stress, we expect  $w_\mathrm{NEDE}(t_*) > 1/3$, which eventually asymptotes to $1/3$ as the condensate decays. Correspondingly, the NEDE plasma subsides quicker than the dominant radiation plasma, effectively shutting down the energy injection. In the case of cold NEDE this is crucial for having a successful phenomenology.

As mentioned before, for the thermal description to be self-consistent, $\psi$ and the gauge field $A$ have to be in thermal equilibrium. This is the case provided $n_A \langle\sigma v\rangle > H$, where $\langle\sigma v\rangle$ is the velocity-averaged cross section for the elastic process $\psi  A \leftrightarrow \psi A$ and $n_A$ is the gauge boson number density. For a non-relativistic gauge boson mass $m_A=g_\mathrm{NEDE} \psi$, we have $n_A \simeq g_A T_d^3 x^{3/2} \mathrm{e}^{-x}/(2 \pi)^{3/2}$, where $x^2= m^2_A/T^2_d \simeq \pi/\gamma \gg1$.  We also used that $\psi \simeq \mu / \sqrt{\lambda}  \simeq g_\mathrm{NEDE} T_d^* / \sqrt{4 \lambda}$ and $T_d^* \simeq T_\circ = 2 \mu / g_\mathrm{NEDE}$ close to a supercooled phase transition. If we substitute the estimate $\langle\sigma v\rangle = c_0 g_\mathrm{NEDE}^4/m_A^2$ (assuming an interaction term $g_\mathrm{NEDE}^2 \psi^2 A^2$), where $c_0$ is a model-dependent factor of order unity, set $g_A=2$ and $H \simeq ( \pi^2/90)^{1/2} \sqrt{3.4}\, T_\mathrm{vis}^2/M_\mathrm{Pl}$, we obtain the lower bound $\lambda >  \sqrt{3.4}\pi^{7/4}  / (4\sqrt{45} )\, T_\mathrm{vis}^* /(c_0\xi_* M_\mathrm{Pl}) \mathrm{e}^{\sqrt{\pi / \gamma}}$. 
We next use \eqref{xi_large_mass} and \eqref{eq:T_d_star_large_mass} to rewrite the bound on $\lambda$ in terms of phenomenological NEDE parameters as
\begin{multline}\label{bound_lambda}
\lambda>4.5 \times 10^{-28}\,\frac{1}{c_0 } \left(\frac{0.4}{\xi_*}\right)  \left( \frac{1+z_*}{5000} \right) \\
\times  \exp{\left[8.3 \left(\frac{ f_\mathrm{NEDE}/[1-f_\mathrm{NEDE}]}{0.1}\right)^{1/2}\left( \frac{0.4}{\xi_*}\right)^2\right]} \, .
\end{multline}
This indeed can be easily satisfied.

\section{Hot NEDE and neutrino mass generation}\label{neutrino_masses}

We argue that the NEDE phase transition can be related to the origin of neutrino masses through the inverse seesaw mechanism~\cite{Abada:2014vea}. For simplicity, we consider the case of a single generation of active left-handed, $\nu_L$, right-handed, $\nu_R$, and sterile, $\nu_s$, neutrinos (the case with three generations is discussed in~\cite{long}). Writing $N \equiv (\nu_L, \nu_R^c, \nu_s)^T$, we will consider the neutrino mass term 
\beq
\mathcal{L}_{\nu} = - \frac{1}{2} N^T C M N + \mathrm{h.c.} \,,
\eeq
where $C$ is the charge conjugation matrix and the mass matrix  $M$ takes the form
\beq\label{massmatrix}
M =\left(\begin{matrix}{}
  0 & d & 0\\
  d & 0 & n \\
 0 & n & m_s
\end{matrix}\right)~.
\eeq
Here, $d = \mathcal{O}(100\, \mathrm{GeV})$ and $n >  \mathcal{O}(\mathrm{TeV})$ are the high-energy entries corresponding to a Dirac mixing of $\nu_R$ with $\nu_L$ and $\nu_s$, respectively.  On the low-energy side, we include a Majorana mass  $\mathrm{eV} < m_s < \mathrm{GeV}$ for $\nu_s$. This mixing matrix then gives rise to a light eigenstate with mass\footnote{We use the subscript ``3'' to indicate that this is the scale of the heaviest of the three mass eigenstates (assuming normal ordering). For more generations, the remaining mass eigenstates, $m_1$ and $m_2$, are further suppressed~\cite{long}.} $m_3 \simeq  m_s  \kappa^2$, where $\kappa = \mathcal{O}(d)/\mathcal{O}(n)  <1$, alongside a pair of heavy pseudo-Dirac fermions with mass $\mathcal{O}(n)$. For example, for $m_s = \mathcal{O}( 100\, \mathrm{eV})$, we require $n =\mathcal{O}( 10 \, \mathrm{TeV})$ (or $\kappa =\mathcal{O}( 0.01)$ equivalently) to account for an active neutrino mass of order of $m_3 =\mathcal{O}( 0.05\, \mathrm{eV})$. If generalized to three generations of neutrinos, this low-energy seesaw mechanism can explain the observed mass spectrum and mixing pattern~\cite{Abada:2014vea}.  

As an aside, it is possible to consider a generalization with a different number of sterile and right-handed neutrinos.  This then allows for the presence of a fourth eV mass eigenstate $\nu_4$ mainly composed of sterile neutrinos. It has been claimed~\cite{Kopp:2013vaa,Boser:2019rta,Dasgupta:2021ies} that the presence of $\nu_4$ can resolve different anomalies seen in accelerator~\cite{Gariazzo:2015rra,Gonzalez-Garcia:2015qrr}, reactor~\cite{Mention:2011rk} and gallium experiments~\cite{Acero:2007su,Giunti:2010zu}, although the role of steriles in resolving these discrepancies is currently debated~\cite{MicroBooNE:2021tya,TheMicroBooNECollaboration:2021cjf,Arguelles:2021meu}. However, if $\nu_4$ equilibrates with the active neutrinos in the early universe, this proposal runs into problems with cosmological bounds on $N_\mathrm{eff}$. It has been argued that this conclusion can be avoided by preventing $\nu_4$ from thermalizing due to a ``secret interaction''~\cite{Hannestad:2013ana,Dasgupta:2013zpn,Archidiacono:2014nda,Archidiacono:2015oma,Archidiacono:2016kkh}. In \cite{long}, we argue that the NEDE field not only gives mass to the neutrinos but also provides the secret interaction to avoid the cosmological bounds.

If we assign lepton number $L=1$ to the sterile neutrino, we notice that the sterile Majorana mass term, $\propto \overline{\nu^c_s} \nu_s$, violates lepton number by two units. Since the lepton symmetry is restored in the limit $m_s \to 0$, it is technically natural to have a small, non-vanishing mass~$m_s$. Once we diagonalize the mass matrix, this protects the light active neutrino mass $m_3$.
The possibility that lepton symmetry is spontaneously broken by $\Psi$ as it acquires a vev in the NEDE phase transition and gives mass to the sterile neutrino follows naturally. If we assign lepton number $L=-2$ to $\Psi$, it allows for a Yukawa interaction term of the form $ \propto g_s \Psi  \, \overline{\nu_s^c} \nu_s$. However,  in order to explain also the origin of the $> \mathrm{TeV}$ Dirac mixing between $\nu_R$ and $\nu_s$, the model must be embedded into a larger dark symmetry group, which is broken in two steps. We will assume that the symmetry group of the dark sector has the form 
\begin{align}\label{G_D}
G_\mathrm{D} \times G_\mathrm{NEDE}
\end{align}
 with charge assignments to allow for the Yukawa couplings
\begin{align}
 \mathcal{L}_\mathrm{Y} &=-g_{\Phi} \Phi  \overline{\nu_R} \nu_s  - \frac{ g_s}{\sqrt{2}} \Psi \overline{\nu_s^c} \nu_s  + g_H \overline{\nu_R} L^T \epsilon H + \mathrm{h.c.}~, \label{Yukawas}
\end{align}
where $L=(\nu_L, e_L)$ is the SM lepton doublet and $H$ the Higgs doublet. We also introduced another dark sector Higgs field $\Phi$ to induce the first breaking.  To be precise, $G_\mathrm{D}$ is broken above the TeV scale, generating the Dirac entry $n = g_\Phi v_\Phi / \sqrt{2}$ as $\Phi \to v_\Phi/\sqrt{2}$. The subsequent electroweak breaking leads to $d = g_H v_H/\sqrt{2}$ with $v_H = 246\, \mathrm{GeV}$, which couples $\nu_R$ and $\nu_L$ . Subsequently, when $G_\mathrm{NEDE}$ is broken at the eV scale,  $\Psi \to v_\Psi / \sqrt{2}$, leading to the Majorana mass $m_s = g_s v_\Psi $. We can in fact relate $m_s$ to  the phenomenological hot NEDE parameters $f_\mathrm{NEDE}$ and $z_*$ when we  use as before that $v_\Psi \simeq g_\mathrm{NEDE} T_d^* / (2 \sqrt{\lambda})$ (again for $\gamma \lesssim 1$ and $\delta_\mathrm{eff} \ll \pi \gamma$) along with \eqref{eq:gamma}, \eqref{eq:delta_large_mass} and~\eqref{eq:T_d_star_large_mass}:
\begin{multline}\label{mass_sterile}
m_s \approx (1.2 \, \mathrm{eV}) \times \frac{1}{\gamma^{1/4}} \frac{g_s}{g_\mathrm{NEDE}}\\ 
\times  \left[\frac{f_{\textrm{NEDE}} /(1-f_{\textrm{NEDE}} )}{0.1}\right]^{1/4} \left[ \frac{1+z_*}{5000}\right]\,
\end{multline}
In particular, for  $ \gamma g_\mathrm{NEDE}^4 =4 \pi \lambda < g_s^4$ the model gives rise to a super-eV sterile as required by the inverse seesaw while allowing for a sizable fraction of NEDE that decays at redshift $z_* \sim 5000$.  

\begin{figure}[t]
    \centering
    \includegraphics[width=0.45\textwidth]{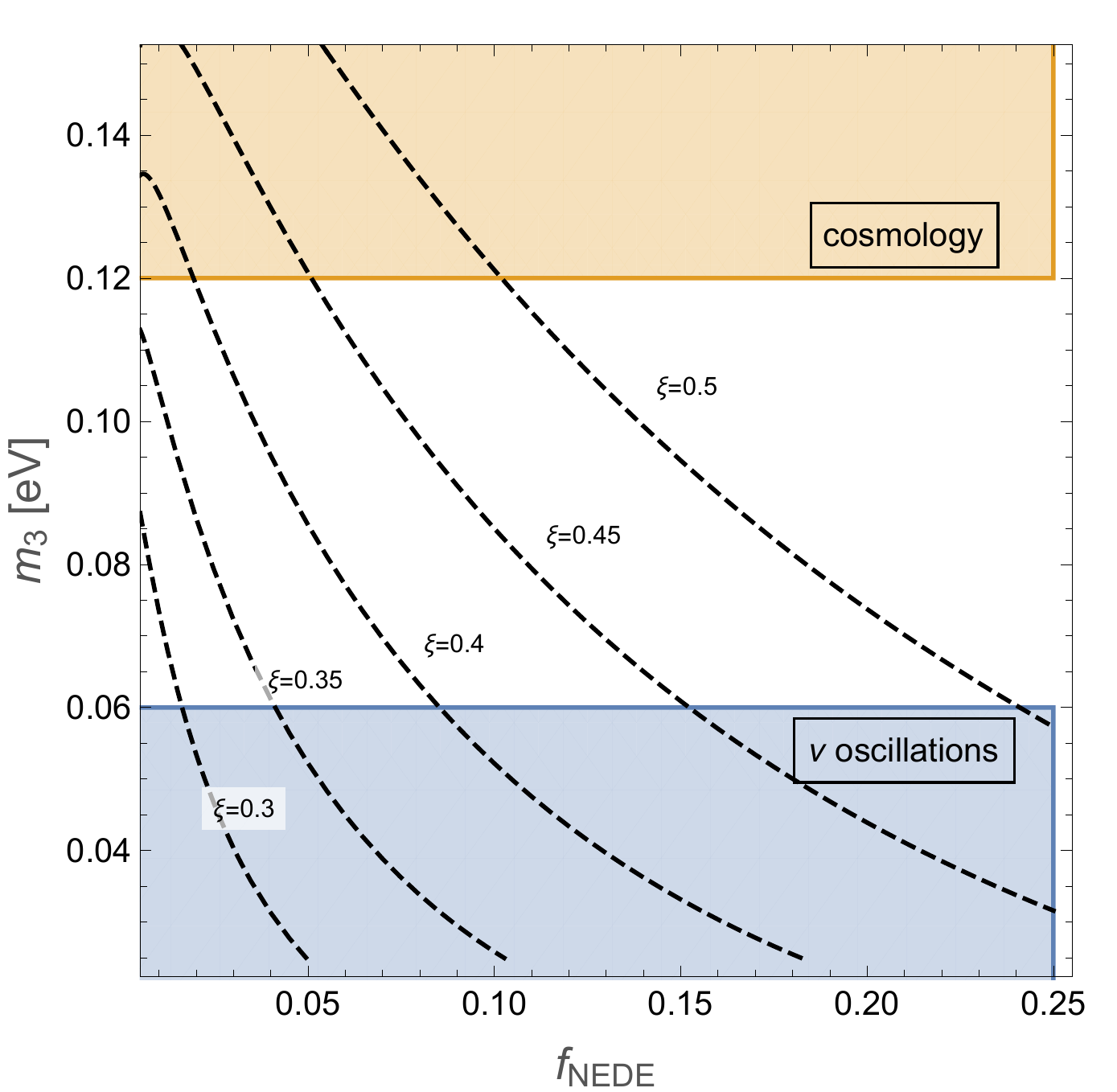}
    \caption{Bounds on the mass of the heaviest neutrino $m_3$ (assuming normal mass ordering). The orange contour is the usual cosmological bound within $\Lambda$CDM~\cite{Planck:2018vyg}, and the blue contour is the lower bound imposed by neutrino oscillations.  The dashed lines represent the upper bound \eqref{bound_m3} arising in hot NEDE for $c_0=1$, $z_*=5000$, and different choices of the dark sector temperature $\xi_*$. Overall, we find that hot NEDE strengthens the existing cosmological bound.}
    \label{fig:m3}
\end{figure}

As we argue here, having a relation between the hot NEDE parameters and $m_s$ makes this model predictive with regard to the neutrino mass spectrum. Due to the sterile-active mixing (mediated by $\nu_R$ and set by the scale $\kappa$), the active neutrinos feel the presence of~$\Psi$, which acts like a secret interaction. This will affect their streaming properties at late times and leads to the bound~\cite{Archidiacono:2013dua} $g_s < 10^{-7} / \kappa^2$ (the factor $\kappa^2$ accounts for the mixing suppression). Substituting this, alongside the lower bound on $\lambda$ in \eqref{bound_lambda}, back into \eqref{mass_sterile} and using $m_3 \simeq  m_s  \kappa^2 $ gives rise to an upper bound on the mass of the heaviest active neutrino:
\begin{multline}\label{bound_m3}
m_3  <  (0.087\, \mathrm{eV} ) \times  c_0^{1/4} \left(\frac{0.4}{\xi_*} \right)^{1/4}\mathrm{e}^{2.1- \frac{1}{\xi^2_*}\left(\frac{f_\mathrm{NEDE}}{1-f_\mathrm{NEDE}}\right)^{1/2} }\\
\times \left[\frac{f_{\textrm{NEDE}} /(1-f_{\textrm{NEDE}} )}{0.1}\right]^{1/4} \left[ \frac{1+z_*}{5000}\right]^{3/4}\,.
\end{multline}
We see that the bound is exponentially sensitive to the value of $f_\mathrm{NEDE} $ and $\xi_*$. As we show in Fig.~\ref{fig:m3}, for  the typical choice $z_* = 5000$, we strengthen the existing cosmological bound~\cite{Planck:2018vyg} $m_3 < \sum_i m_i \leq 0.12$ (orange region), provided the model-dependent factor $c_0$ takes on values of order unity (a more precise statement will require a detailed calculation within a particular dark sector model). In fact, we strengthen it so much that we get close to the lower bound $m_3 > 0.06 \mathrm{eV}$ arising from oscillation data (blue region). This observation can be used to obtain a lower bound on $\xi_*$. In the somewhat forgiving scenario where $c_0 =10$, $z_*=20000$ (the largest value which in principle still allows to address the Hubble tension),  we find $\xi_* > 0.3$ if we want $f_\mathrm{NEDE} =0.1 $. At the same time, we do not tolerate $\xi_* > 0.5$ as this would give a sizable contribution to the equivalent number of neutrino species~\cite{Buen-Abad:2015ova} $\Delta N_\mathrm{eff} = \frac{4}{7}(\frac{11}{4})^{4/3}  g_\mathrm{rel,d}\xi^4 >  0.13$ (corresponding to a dark radiation rather than early dark energy scenario), leaving us with the narrow range
\begin{align}\label{range_xi}
0.3 < \xi_* < 0.5\,.
\end{align}
A precise fit of hot NEDE to CMB data will provide a falsifiable test of this prediction. 

To summarize, we showed that the hot NEDE implementation of the inverse seesaw is falsifiable through \eqref{bound_m3} and \eqref{range_xi} once we constrain   $\xi_*$, $f_\mathrm{NEDE}$ and $z_*$ through a fit to cosmological data. At the same time, \eqref{range_xi} allows us to constrain different microphysical implementations of hot NEDE. This will be exemplified in the next section when we discuss the DEW model.

\section{Dark Electroweak Model}\label{sec:DEW}
We start by reviewing and motivating the DEW model as introduced in \cite{long}.
Since the Majorana mass breaks lepton number, the minimal choice is to identify $G_\mathrm{NEDE}$ in \eqref{G_D} with the global $\mathrm{U(1)}_\mathrm{L}$ lepton number symmetry. Since its breaking occurs in the dark sector, the corresponding massless goldstone boson, the \textit{majoron}, will only contribute to the (subdominant) dark radiation fluid.  Of course, in reality, the Goldstone will not be exactly massless as global symmetries are believed to be broken by quantum gravity effects~\cite{Kallosh:1995hi}. The possibility of including a very light majoron is discussed in~\cite{long}.

A similar minimal choice would then be to identify $G_\mathrm{D} $ with a new $\mathrm{U(1)_D}$ gauge symmetry under which $\nu_s$ and $\Phi$ would carry opposite charges to allow for the Yukawa in \eqref{Yukawas}. However, this choice is too minimal to allow for a strong first-order NEDE phase transition because after the high-energy breaking, it lacks the light bosonic degrees of freedom needed to induce the thermal barrier in \eqref{eq:effective_T_pot_low_T} around the eV temperature scale.\footnote{It is possible that this can be fixed by adding higher dimensional non-renormalizable operators to the potential or by gauging the $\mathrm{U(1)_{L}}$ symmetry.} 

As the next-to-minimal option, we will assume that the dark sector transforms in a dark copy of the electroweak group. In this DEW model, we, therefore, have $G_\mathrm{D}=\mathrm{SU(2)_D}\times \mathrm{U(1)_{Y_\mathrm{D}}}$, where $Y_\mathrm{D}$ refers to a dark hypercharge. The corresponding gauge couplings are $g_d \simeq g'_d$. Now, we assume that the sterile $S=(\nu_s, S_-)^T$ and the Higgs $\Phi = (\Phi_+, \Phi_0)^T $ transform as a doublet with $Y_{\mathrm{D},S}= - Y_{\mathrm{D},\Phi} = -1 $, while the NEDE field is promoted to a triplet $\Psi = (\Psi_1,\Psi_2,\Psi_3)^T$ with $Y_{\mathrm{D},\Psi} = 2$. The SM fields and $\nu_R$, on the other hand, transform as singlets with $Y_{\mathrm{D}} = 0$. This generalizes \eqref{Yukawas} to 
\begin{multline}
 \mathcal{L}_\mathrm{Y} =-g_{\Phi} \overline{\nu_R} S^T \epsilon \Phi  - \frac{ g_s}{2 } \overline{S^c} \epsilon \Delta S + g_H \overline{\nu_R} L^T \epsilon H + \mathrm{h.c.}~, \label{Yukawas2}
\end{multline}
where $\Delta = \Psi \cdot \tau $  with $\tau = (\tau_1,\tau_2 ,\tau_3)$ the Pauli matrices. This can be understood as a dark sector version of the Gelmini-Roncadelli model with renormalizable potential~\cite{Gelmini:1980re}
\begin{multline}\label{fullpotential}
V= a \Phi^\dagger \Phi + c \left( \Phi^\dagger \Phi \right)^2  - \frac{ \mu^2}{2} \mathrm{Tr} \left( \Delta^\dagger \Delta\right) + \frac{\lambda}{4} \left[ \mathrm{Tr} \left( \Delta^{\dagger}\Delta \right) \right]^2\\
 + \frac{e-h}{2} \Phi^\dagger \Phi  \mathrm{Tr} \left( \Delta^{\dagger}\Delta \right) + h \Phi^\dagger \Delta^\dagger \Delta \Phi +\frac{f}{4} \mathrm{Tr}\left( \Delta^\dagger \Delta^\dagger\right)   \mathrm{Tr}\left( \Delta \Delta\right) \,.
\end{multline}
The high-energy breaking is triggered when $\Phi $ picks up its vev, $\Phi \to (0, v_\Phi/\sqrt{2})^T $. However, the breaking direction is chosen such that it preserves a dark electromagnetism group, $G_\mathrm{D} \to \mathrm{U(1)}_\mathrm{DEM}$. Accordingly, the components of $\Psi$ can be decomposed into neutral, single-charged and double-charged states as $\Psi_{\smash{{}^1_2}} =  \pm 1/\sqrt{2} (\Psi_0 \pm \Psi_{++})$ and $\Psi_3 = \Psi_+$. The NEDE scalar is then identified with the neutral component $\Psi_0$. During the NEDE phase transition $\psi \equiv \sqrt{2} |\Psi_0|\to v_\Psi$, which implies a breaking of the global $\mathrm{U(1)}_\mathrm{L}$ by two units and, as shown before, gives mass to the active neutrinos.
From \eqref{fullpotential} we derive the vacuum  condition
\begin{subequations}
\label{vev_system}
\begin{align}
a + c v_\Phi^2  + \frac{1}{2} \left( e-h \right) v_\Psi^2 &= 0 \,,\\
-\mu^2 + \lambda v_\Psi^2  + \frac{1}{2} \left( e-h \right) v_\Phi^2  &= 0 \,.
\end{align}
\end{subequations}
We further assume a separation of scales where $v_\Psi \ll v_\Phi$. 
This is compatible with \eqref{vev_system} without invoking a finetuning if both equations decouple, which, in turn, requires $e,h \lesssim \lambda v_\Psi^2 / v_\Phi^2 \ll 1$. This condition is technically natural for a sufficiently small dark gauge coupling obeying~\cite{long} $g_d^2 \lesssim \mu / v_\Phi $ and $g_d^4 \lesssim \lambda $.
 In other words, the gauge couplings have to be hierarchically small. This makes thermal corrections arising from the gauge bosons too small to achieve supercooling ($\gamma \lesssim 1$). However, we can still make contact with the hot NEDE discussion when noting that for $g_d^2 \ll f$, the dominant corrections are caused by the coupling with $\Psi_{++}$. This motivates the identification~\cite{long} $g_\mathrm{NEDE}^2 \sim f$ and makes the analysis of Sec.~\ref{sec:hotNEDE} applicable, although a detailed computation of the temperature corrected potential in the DEW model is still outstanding. 

The spectrum of the low-energy theory contains the massless majoron, the NEDE scalar fluctuation $\delta \psi \equiv \psi - v_\Psi$ with mass $m_\Psi^2 \simeq 2 \lambda v_\Psi^2 \simeq 2 \mu^2$, the single-charged component $\Psi_+$ with mass $m_{\Psi_+} < m_\psi$, and the double-charged component $\Psi_{++}$ with mass $m_{\Psi_{++}}^2 \simeq 2 f v_\Psi^2 \gg m_\psi$ (assuming $\gamma \lesssim 1$ or $f^2 \gtrsim 4 \pi \lambda$ equivalently). Along with three massive and one massless gauge boson, this corresponds to 17 bosonic degrees of freedom. If $g_\mathrm{rel,d} (\leq 17)$ of them are relativistic and in thermal equilibrium with the dark radiation plasma, this amounts to a contribution~\cite{Buen-Abad:2015ova} $\Delta N_\mathrm{eff} = \frac{4}{7}(\frac{11}{4})^{4/3}  g_\mathrm{rel,d}\xi^4$ to the effective number of neutrino species. For it to be negligible, we demand conservatively $\Delta N_\mathrm{eff} < 0.1$ (in agreement with CMB bounds~\cite{Planck:2018vyg}), which translates to an upper bound $\xi \lesssim 0.46/g_\mathrm{rel,d}^{1/4}$. It is in agreement with the bound in \eqref{range_xi} if $g_\mathrm{rel,d} \leq 6$. For example, a scenario with $g_\mathrm{rel,d} = 6$ could correspond to the case where the gauge bosons are all decoupled from the thermal plasma, which is compatible with our assumption on the gauge coupling, $g^2_d \ll g^2_\mathrm{NEDE} \sim f $. Due to \eqref{xi_large_mass}, it is also compatible with $f_\mathrm{NEDE} = 10\%$ within the supercooled regime corresponding to $\gamma \lesssim 0.08 / g_\mathrm{rel,d} $. In summary, the DEW model exemplifies that the hot NEDE field can be connected to the origin of neutrino masses and the spontaneous breaking of lepton number conservation while being compatible with the phenomenological bounds in \eqref{bound_m3} and \eqref{range_xi}.

Finally, for the inverse seesaw mechanism to work, we need $v_\Phi > \mathrm{TeV}$ to create a moderate hierarchy between $d$ and $n$ in \eqref{massmatrix}. This offers a natural way to also create the {Dirac} mass $M_\chi$ of a super-TeV DM candidate $\chi = \chi_L + \chi_R$, where $\chi_R$ is a a singlet and $\chi_L$ is the charged component of an $\mathrm{SU(2)_D}$ doublet $X = (\chi_0,\chi_L)^T$. Explicitly, we can now add to \eqref{Yukawas2} the Yukawa coupling term
\begin{align}
\mathcal{L}_\mathrm{Y} \supset - g_\chi \overline{\chi_R} X \Phi + \mathrm{h.c.}~,
\end{align}
giving rise to $M_\chi = g_\chi v_\phi /\sqrt{2} $, where $g_\chi$ is the Yukawa coupling parameter.
This DM candidate can be produced through gravitational interactions at high energies via the freeze-in mechanism, as argued in the Planckian interacting DM proposal in \cite{Garny:2015sjg,Garny:2017kha}. Moreover, as discussed in \cite{long}, as it is charged under $\mathrm{U(1)}_\mathrm{DEM}$, it introduces DM-DR interactions with possible applications to the LSS tension~\cite{Lesgourgues:2015wza,Buen-Abad:2017gxg,Archidiacono:2019wdp}. The neutral component $\chi_0$, on the other hand, will make a contribution to the dark sector radiation plasma.

\section{Observational Signatures}\label{sec:pheno}

The proposed model leads to signatures in both particle physics experiments and cosmological observations. While a quantitative discussion goes beyond this letter's scope, in the following, we provide a qualitative summary of how different signatures can be used to constrain the model.
\begin{itemize}
\item \textbf{CMB:} Since the phase transition happens during the CMB epoch, it affects the acoustic oscillations in the primordial plasma. The main effect is a reduction of the sound horizon, which needs to be balanced by an increase in $H_0$, resolving the Hubble tension. Beyond that main effect, NEDE will also lead to characteristic changes in the temperature and polarization power spectrum (for a detailed discussion, see Sec.~IIIE in \cite{Niedermann:2020dwg}) that can be searched for in future high-multipole polarization and temperature data. 

\item \textbf{LSS:} NEDE leads to an excess decay of the gravitational potential, which needs to be compensated by an increased DM density $\omega_\mathrm{CDM}$. This, in turn, affects and typically enhances the matter power spectrum. At the same time, the microscopic scenario proposed here will introduce interactions between the dark sector radiation plasma and NEDE ($\psi$) or DM ($\chi$). The latter type of interaction is known to suppress the small-scale power spectrum~\cite{Archidiacono:2019wdp}. As a result, full-shape LSS data will play a crucial role in constraining different microscopic scenarios.   

\item \textbf{Lepton flavor violation:} The inverse seesaw scenario will generically lead to a non-unitarity of the PMNS matrix, which will modify the vertex $W\ell\nu$ (with $\ell = e,\mu, \tau$). Due to the associated charged lepton flavor violation, this will also manifest itself through processes such as $\mu \to e \gamma$~\cite{Abada:2014vea}. The strength of this effect will be controlled by the parameter $\kappa$. Similarly, this proposal opens new leptonic Higgs decay channels such as $H \to e \bar{\mu}$~\cite{Arganda:2014dta} (for other lepton flavor violating processes that can also be affected see~\cite{Abada:2012mc,Abada:2013aba}). As recently claimed in \cite{Abada:2012mc,Blennow:2022yfm}, these modified interactions have also the potential to resolve the $R_K$ and $R_{K^*}$ anomalies or the W-boson mass anomaly.

\item \textbf{Neutrino sector:}  First, of course, the mass entries in \eqref{massmatrix} have to be chosen such that they can accommodate the observed neutrino oscillation data. Another bound arises from the neutrinoless double beta decay, the strength of which is controlled by the Majorana mass parameter $m_s$. Moreover,  the non-unitarity of the PMNS matrix also introduces non-standard neutrino interactions with SM fermions (for constraints see~\cite{Antusch:2008tz}). Finally, due to the neutrino mass mixing, the second term in \eqref{Yukawas} will induce invisible decays of the light neutrino mass eigenstates to our NEDE scalar $\psi$.  These processes are constrained by CMB and LSS data~\cite{Archidiacono:2013dua,Barenboim:2020vrr}. 

\item \textbf{Gravitational waves:} A first-order phase transition leads to a stochastic background of gravitational waves. As the NEDE transition occurs at comparatively low energies, the peak frequency of the corresponding spectrum is not probed by current experiments. However, as argued in \cite{Niedermann:2020dwg}, the high-frequency tail of the spectrum can overlap with the peak sensitivity of future pulsar-timing arrays, provided the phase transition is sufficiently slow.

\end{itemize}

\section{Conclusions}

Hot NEDE relies on thermal corrections induced within a dark sector to trigger a supercooled first-order phase transition around the $\mathrm{eV}$ scale~\cite{long}. The associated false vacuum energy doubles as an early dark energy component and has the potential to resolve the Hubble tension. The same transition can dynamically create the super-eV Majorana mass of a set of sterile neutrinos by spontaneously breaking a global lepton number symmetry. Building on the inverse seesaw mechanism, which further introduces super-$\mathrm{TeV}$  Dirac couplings with new right-handed neutrinos, this can explain the observed masses and mixing angles of the active neutrinos. 

In this work, after reviewing the hot NEDE framework alongside the inverse seesaw mechanism, we searched for consistency relations between hot NEDE and the active neutrino sector. We found one in the form of a rather stringent upper bound on the mass of the heaviest active neutrino $m_3$ [see Eq.~\eqref{bound_m3}].
It is most sensitive to the amount of NEDE, $f_\mathrm{NEDE}$, and the dark sector temperature at the moment of the phase transition, $\xi_*$.  In particular, having $f_\mathrm{NEDE} > 0.1$, as required for resolving the Hubble tension, together with a cold dark sector  $\xi_* < 0.5$, as preferred by a supercooled transition, strengthens existing cosmological bounds.  In other words, hot NEDE provides us with a falsifiable particle physics prediction that does not exist in the original inverse seesaw implementation and originates from the fact that the model attempts to explain both the Hubble tension and the origin of neutrino masses. 
Moreover, being compatible with the existing lower bound on the neutrino masses arising from oscillation data imposes a lower bound on $\xi_*$, leaving us with $0.3 < \xi_* < 0.5$, a prediction that can be tested by precise fits of hot NEDE to CMB. This finite range then informs model building within the hot NEDE framework. 
We showcase this within the DEW model, which makes a particular choice for the dark sector gauge group, $G_\mathrm{D}=\mathrm{SU(2)}_\mathrm{D} \times \mathrm{U(1)_\mathrm{Y_D}} $, and identifies the NEDE field as the neutral component of an $\mathrm{SU(2)}_\mathrm{D}$  triplet. As a result, we find that the model can fulfill the bounds provided the dark gauge sector is sufficiently weakly coupled.

Finally, we highlighted the observational signatures of this scenario that can be looked for in cosmological (CMB, BAO, LSS, gravitational waves) and particle physics data (flavor violating processes, Higgs decay, non-unitarity of PMNS matrix, non-standard neutrino interactions, neutrinoless double beta decay). 

In conclusion, we have found two falsifiable predictions of the hot NEDE framework. A stringent upper bound on the heaviest active neutrino mass $m_3$, and a narrow allowed range for the dark sector temperature $\xi_* $.
In our future work, in addition to testing the prediction of the dark sector temperature by detailed fits to CMB data, we will investigate if the lepton symmetry breaking in the NEDE phase transition can shed light on accelerator anomalies such as the $(g-2)_{\mu}$ anomalous magnetic moment of the muon, the $R_K$ and $R_{K^*}$ anomalies or the W-boson mass anomaly. To be more specific, the idea is to look at deviations from unitarity in the PMNS matrix as they naturally arise in the inverse seesaw mechanism~\cite{Abada:2012mc,DelleRose:2019ukt,DelleRose:2020oaa,Blennow:2022yfm} and/or introduce mixing effects between the visible and dark sector~\cite{Zhang:2022nnh}.

\section*{Acknowledgments}
We would like to thank Edmund Copeland and Steen Hannestad for their useful comments on the draft.
This work is supported by Villum Fonden grant 13384 and Independent Research Fund Denmark grant 0135-00378B.



  \bibliographystyle{elsarticle-num} 
  \bibliography{Hot_NEDE}





\end{document}